%% file: Main.tex
\documentclass[
    oneside,
    fontsize=10pt,leqno]{article}
\usepackage[T1]{fontenc}
\usepackage[english]{babel}
\usepackage{enumitem}
\usepackage{amsfonts,amssymb,amsmath,amsthm}
\usepackage{comment}
\usepackage{tikz}
\usepackage{appendix}
\usepackage{natbib}

\input{jite_regular}
\input{jite_style01}

\usepackage{booktabs} 
\usepackage{caption}
\usepackage{adjustbox}
\usepackage{relsize}
\usepackage{url}
\usepackage{epstopdf}
\usepackage{graphics,graphicx}
\usepackage{balance}
\usepackage{array}
\usepackage{pifont}
\usepackage{multirow}
\usepackage{xcolor}
\usepackage{footmisc}
\usepackage[normalem]{ulem}
\usepackage{paralist, tabularx}

\newcommand{\buybox}[0]{Buy Box}
\newcommand{\new}[1]{\textcolor{black}{#1}}

\begin{document}

\title{Antitrust, Amazon, and Algorithmic Auditing}

\author{Abhisek Dash, Abhijnan Chakraborty, Saptarshi Ghosh, Animesh Mukherjee, Jens Frankenreiter, Stefan Bechtold and Krishna P. Gummadi\thanks{
Dash \& Gummadi (corresponding authors): Max Planck Institute for Software Systems, Saarbrücken, Germany; Chakraborty: Indian Institute of Technology Delhi, India; Ghosh \& Mukherjee: Indian Institute of Technology Kharagpur, India; Frankenreiter: Washington University in St. Louis, USA; Bechtold: ETH Zurich, Switzerland. This research is supported in part by a European Research Council (ERC) Advanced Grant for the project ``Foundations for Fair Social Computing'', funded under the European Union's Horizon 2020 Framework Program (grant agreement no. 789373), and by a grant from the Max Planck Society through a Max Planck Partner Group at IIT Kharagpur. A. Dash was supported by a fellowship from Tata Consultancy Services for major duration of this research. Excellent research assistance by Emma Deering and Simona Ramensperger is gratefully acknowledged.}}

\maketitle

\begin{abstract}
In digital markets, antitrust law and special regulations aim to ensure that markets remain competitive despite the dominating role that digital platforms play today in everyone's life. Unlike traditional markets, market participant behavior is easily observable in these markets. We present a series of empirical investigations into the extent to which Amazon engages in practices that are typically described as self-preferencing. We discuss how the computer science tools used in this paper can be used in a regulatory environment that is based on algorithmic auditing and requires regulating digital markets at scale.



\JEL{K21, L40, L42}
\end{abstract}

\section{Introduction}\label{sec:Introduction}
\input{1_Introduction}

\section{Amazon and Its Marketplace}\label{sec:Amazon}
\input{2_Amazon}

\section{Amazon's Vertical Integration and Antitrust Laws}\label{sec:antitrust}
\input{3_Antitrust}

\section{Amazon and Preferential Treatment}\label{sec:preftreatment}
\input{4_Preferential_Treatment}

\section{Discussion}
\input{5_Discussion}

\section{Conclusion}
\input{6_Conclusion}
	
\bibliography{Main}

\end{document}

%% file: jite_regular.tex
\long\def\thanks#1{\begingroup\def\thefootnote{\fnsymbol{footnote}}\footnote[1]{#1}\endgroup}
\makeatletter
\def\maketitle{
   \null%
    \vskip 1cm
    {\centering{
    {\Large \@title\par}%
    \vskip 0.4cm%
    {\normalsize by} %
   \vskip 0.3cm%
    {\normalsize \@author}%
    \vskip 0.9cm%
    {\small \@date}%
    \vskip 0.4cm}%
             }}
\makeatother

%% file: jite_style01.tex
\usepackage{geometry}
\usepackage[defaultlines=2,all]{nowidow}

\usepackage{lscape}

\renewcommand{\normalsize}{\fontsize{10}{11.6}\selectfont}
\renewcommand{\footnotesize}{\fontsize{8}{9}\selectfont}

\makeatletter
\renewenvironment{quote}
               {\list{}{\footnotesize
                        \leftmargin=0.4cm
                        \rightmargin=0.4cm
                        \parsep=6pt}
                \item\relax}
               {\endlist}
\makeatother

\renewenvironment{abstract}{\begin{quote}}{\end{quote}}
\newcommand{\Keywords}[1]{\textit{Keywords:} #1}
\newcommand{\JEL}[1]{\textit{JEL classification code:} #1}

\usepackage{titlesec}
\titleformat*{\section}{\centering\normalsize\itshape}
\titlespacing*{\section}{0pt}{20pt}{6pt}
\titleformat*{\subsection}{\normalsize\itshape}
\titlespacing*{\subsection}{0pt}{12pt}{6pt}
\titleformat*{\subsubsection}{\normalsize\rmfamily}
\titlespacing*{\subsubsection}{0pt}{12pt}{6pt}
\titleformat*{\paragraph}{\normalsize\itshape}
\titlespacing*{\paragraph}{0pt}{8pt}{10pt}

\renewcommand{\appendix}{\par
  \pdfbookmark[1]{Appendix}{appendix}
  \setcounter{subsection}{0}
  \renewcommand{\thesubsection}{A.\arabic{subsection}}
  \setcounter{subsubsection}{0}
  \renewcommand{\thesubsubsection}{A.\arabic{subsection}.\arabic{subsubsection}}
  \setcounter{equation}{0} \renewcommand{\theequation}{A\arabic{equation}}
  \setcounter{table}{0}
  \renewcommand{\thetable}{A\arabic{table}}
  \setcounter{figure}{0}
  \renewcommand{\thefigure}{A\arabic{figure}}}

\setlist[itemize]{label=$-$,topsep=6pt,
parsep=0pt,itemsep=1pt,leftmargin=*,align=left}

\setlist[enumerate]{topsep=6pt,parsep=0pt,itemsep=1pt,
    leftmargin=*,align=left,widest=l)}

\usepackage[breaklinks,hidelinks]{hyperref}
\usepackage[noabbrev]{cleveref} 

\usepackage{hyphenat}
\makeatletter
\g@addto@macro\normalsize{%
  \setlength\abovedisplayskip{8pt}
  \setlength\belowdisplayskip{8pt}
  \setlength\abovedisplayshortskip{4pt}
  \setlength\belowdisplayshortskip{4pt}
}
\makeatother

\newtheoremstyle{JITEth}
{8pt}{8pt}{\itshape}{ }{\scshape}{ }{ }
{\thmname{#1} \thmnumber{\textnormal{#2}}\thmnote{\ (#3)}}
\theoremstyle{JITEth}

\usepackage{ushort}

\allowdisplaybreaks[3]

\usepackage{array}
\usepackage[font=footnotesize,labelfont=it,labelsep=newline,justification=centering]{caption}
\usepackage[para]{threeparttable}
\usepackage{multicol,multirow}
\usepackage{footnote}\makesavenoteenv{tabular}
\usepackage{ucs}

\usepackage[FIGTOPCAP,center,footnotesize]{subfigure}
\usepackage[figuresright]{rotating}

\usepackage[normalem]{ulem}

\newenvironment{2Addresses}
{\bigbreak\parindent=0pt\setlength{\tabcolsep}{0pt}\small
\begin{tabular}{p{5.5cm}@{\hspace{1cm}} p{5.5cm}}}
{\end{tabular}}

%% file: 1_Introduction.tex
In digital markets, a small number of digital platforms often mediate transactions between a large number of firms on the one side and millions of consumers on the other side. The platforms typically benefit from network effects that enable them to navigate in two-sided markets in a profit-maximizing way. Over the last years, policymakers around the world have discussed the extent to which such platforms should be regulated in order to preserve competition in digital markets, maintain a free discourse on the Internet, protect the privacy interests of consumers, and secure the intellectual property rights of authors and inventors. 

Regulating digital platforms can be challenging. One reason is the fact that the Internet has torn down national borders and facilitated communication and trade across the globe. While, in principle, the transnational scope of platform activities seems to require a global response, international coordination mechanisms have often proven ineffective. Against this background, individual countries and regions have increasingly enacted legal regimes to govern the digital economy, even though such regulatory activity raises challenging questions about regulatory ``spillovers'' \citep{bradford2012,bradford2020,frankenreiter2022,peukert2022,marottawurgler2024}.

Maybe even more importantly, regulators face a problem of scale. Digital platforms facilitate billions of transactions between firms and consumers. Google, for example, processes billion searches and billion advertisement auctions every day; Facebook connects over 3 billion people every month \citep{Statista_Facebook_2023}; and users stream almost 1 billion hours of video on YouTube every day \citep{Cloud_YouTube_2023}. There are about 3.5 million mobile apps on the Google Play store, about 1.6 million apps on the Apple App store, and about 200 million active websites \citep{tambe2024,statista2023}.
In this environment, keeping track of the actions of digital platform can be challenging. While digitization has led to a drastic decrease in transaction costs and a drastic increase in scale efficiencies for companies, public authorities have not benefited from similar efficiency gains when enforcing digital regulations against these companies. 

Besides, various characteristic features of digital platforms pose particular challenges for regulators endeavoring to prevent harmful business practices. Platforms do not only establish the rules governing user interactions. They also design the underlying architecture facilitating these interactions. As Lawrence Lessig famously proclaimed about 25 years ago: ``Code is Law'' \citep{lessig1999}. On digital platforms, the ``code'' governing the platforms is written by private companies, not by the legislator. If a regulator wants to influence digital platforms, it has to deal with these privately created ``codes'' and often resort to indirect ways of regulation.

From the perspective of individual users, digital platforms serve as both gateways to information and architects of their interaction spaces. Digital platforms wield considerable influence over what is commonly termed 
\textit{\new{choice architectures}} \citep{thaler2008nudge}, the environments in which users navigate decision-making processes. Extensive research in psychology and related fields has demonstrated that the ability to manipulate choice architectures can profoundly shape user behavior and outcomes \citep{thaler2008nudge}. Digital platforms may use technical choice architectures expressed in algorithms to shape consumer preferences, search behavior and purchase patterns.

These features make it challenging for regulators to tame digital platforms. For one, designing choice architectures lies at the heart of platforms' core functions. Choice architectures exist primarily to facilitate interactions among platform users, and a platform's commercial success often hinges on the effectiveness of these choice architectures in engaging and retaining users. Regulators must therefore carefully consider interventions in platform design and operation, as heavy-handed approaches could impede innovation and hinder platforms' agility in responding to shifting user preferences and market dynamics.

For another, assessing whether a platform utilizes a specific choice architecture to advance its own interests in improper ways poses a considerable challenge in its own right, a topic we explore in detail throughout this article. As we will demonstrate, comparing user interactions within existing choice architectures to hypothetical ``neutral'' alternatives is often impractical, as such alternatives might simply not exist. In the context of digital marketplaces, for instance, what is the benchmark against which to assess the platform's decision to include, among the information displayed to consumers, the total number of transactions a seller has completed--a feature that may favor larger sellers, particularly those associated with the platform provider? And even when a comparison between an existing choice architecture and a ``neutral'' alternative is feasible, distinguishing between design features serving legitimate business objectives and those potentially serving improper 
goals is often challenging. For example, it is not always obvious whether a digital marketplace's choice architecture that appears to nudge customers towards products affiliated with the platform provider is mainly the result of a discrimination of independent competitors or whether it is a response to preferences among consumers for such products, thereby enhancing the overall customer experience.

In this context, a main challenge is that users' expressed preferences can be shaped by and fluctuate with the choice architectures in which users express and act upon their preferences. Put simply, a way to determine the ``true'' preferences of consumers might not always be available, as their preferences may evolve in response to the design of the platforms they interact with. As a consequence of this endogenoneity, assessing whether a design feature aligns with these preferences can be elusive. 

A related aspect is the pervasive role of algorithms in mediating interactions on online platforms. These algorithms dictate which information is presented to users and can significantly influence their action space at any given moment. Digital platforms employ a variety of algorithmic systems to facilitate interactions among their diverse user groups. Posts visible to users on social media platforms, the price offered to drivers accepting rides on a ride-sharing platform, or the products displayed to shoppers on Amazon are all determined by such complex computer programs. Usually, digital platforms keep the precise nature of the factors that influence the algorithms' decisions, as well as the way they are combined, a closely guarded secrets. This opacity renders it challenging for regulators to understand the factors that shape user experiences and behaviors on these platforms, hindering their ability to effectively assess and address potential risks or harms.

Despite these challenges, the digital nature of the transactions mediated by digital platforms also offers new avenues for regulatory enforcement. Compared to traditional markets, market participant behavior is often easier to observe in these markets. Many digital marketplaces can be accessed by everyone, and legislators are increasingly establishing data access rights for regulators, researchers and the wider public that offer opportunities to generate insights into the business models of these platforms. 

That said, obtaining information about business models is usually not trivial. As a first step, a regulator or researcher seeking to assess the operations of a digital platform with the help of digital tools would typically try to access the platform through a regular web interface or engage in automated scraping of the platform's content. However, interfaces of digital platforms have become more interactive, oftentimes rendering simple web access insufficient to capture the inner workings of a digital platform. Digital platforms and markets enabled by these platforms are growing more complex, making it harder to gain a proper understanding of platforms' business practices. 

In response, empirically oriented computer science research has developed various approaches to audit digital platforms over the last few years. \cite{sandvig2014auditing} distinguish between ``code audit\new{[s]}'' (where researchers directly inspect a copy of the relevant algorithm), ``noninvasive user audit\new{[s]}'' (where users interacting with the platform answer questions about what they did online), ``scraping audits'' (where researchers query a platform repeatedly, observing results), ``sock puppet audit\new{[s]}'' (where researchers use computer programs to impersonate users), and ``crowdsourced audit\new{[s]}'' (where researchers employ users as testers). This development points towards a future of algorithmic auditing, where potential adverse effects of algorithmic systems are investigated and evaluated using algorithmic auditing tools \citep{sandvig2014auditing,bandy2021,metaxa2021}. ``The Answer \textit{to} the Machine is \textit{in} the Machine", as Charles Clark pronounced in the 1990's \citep{clark1996}: As digital platforms are increasingly relying of algorithmic systems to govern their digital platforms, regulators will have to resort increasingly to algorithmic systems to audit compliance on these platforms.

This article explores some of the opportunities and challenges of algorithmic enforcement, using Amazon as an example. Recently, Amazon's business policies and deployed algorithmic systems have come under intense scrutiny due to Amazon's vertical integration, or in other words, its involvement in multiple stages of the production and distribution process for its products and services. In this article, we present a series of empirical investigations into the extent to which Amazon engages in practices that are typically described as \new{\textit{preferential treatment or self-preferencing}.} 
In these investigations, we analyze different features of various choice architectures that Amazon presents to its customers. We investigate how the service decides the offer that is featured in its Buy Box (now called ``Featured Offer''); the design feature of its page listing various offers for the same product; how it selects products offered to customers that use the conversational product search technology available through its virtual assistant technology Alexa; and how it displays results in its related-item recommendation system. 

The purpose of this analysis is not to determine whether all or some of the practices we observe constitute violations of antitrust laws or specialized regimes such as the European Union's Digital Markets Act. Conducting a full legal analysis of Amazon's business practices is beyond the scope of this paper. Our aim is to demonstrate how empirical methods originating in computer science can empower policymakers, regulators, and researchers to track digital platform behavior at a granular level, thereby generating empirical evidence that could contribute to discussions on how to effectively regulate digital markets. Concurrently, our discussion underscores how various features of digital platforms can pose challenges for researchers and regulators in deducing anti-competitive intent or impact from observed patterns of platform behavior. Specifically, the case studies presented highlight the considerable challenges involved in assessing whether a platform utilizes a specific choice architecture to advance its own interests in improper ways.

Our article proceeds as follows. Section 2 introduces the design of Amazon's digital marketplace and several choice architectures governing this marketplace. Section 3 provides an overview of how Amazon's vertical integration relates to antitrust and digital market regulation doctrines in the European Union and the United States. 4 Section presents our empirical observations on four distinct types of choice architectures: Buy Box, listing pages, Alex\new{a} search, and recommendation systems. Section 5 discusses our findings. Section 6 concludes.

%% file: 2_Amazon.tex
Over the last years, Amazon has emerged as the largest online e-commerce platform by connecting millions of sellers across the world to billions of customers \citep{Saqib023Most}. On the Amazon platform, the different stakeholders can be primarily classified into three categories: (a)~sellers on the platform \new{--} they rely on Amazon to sell their products to consumers; (b)~customers of the platform \new{--} they rely on Amazon for fulfilling their purchase needs; and (c)~the platform organization itself \new{--} it mediates the interaction between the sellers and costumer.

Due to the enormity and scale at which Amazon operates, Amazon can facilitate these interactions only by relying on mass-scale automation and on countless deliberate design choices that determine how sellers can present their products to customers, and how customers express their preferences in the face of these results. These design choices result in 
choice architectures \citep{johnson2012beyond, mota2020desiderata}. Originating from debates about behavioral policy-making, the term \textit{choice architecture} refers to a deliberate organizing of context in which people make decisions \citep[3]{thaler2008nudge}. As in the behavioral policy making context, choice architectures on digital platforms are the result of conscious design decisions that should be subject to scrutiny by policy makers and regulators. As indicated in the introduction, digital platforms are environments in which the platform provider has full control over the platform's choice architectures. As a result, if a digital platform raises competition problems, these problems can often be traced back to choice architecture decisions the platform provider made.

In Section \ref{sec:preftreatment}, we investigate several choice architectures under the control of Amazon. First, when multiple sellers offer to sell the same product on Amazon, one of the sellers is selected by default by the deployed \textit{\buybox{} algorithm} to feature on the product page \citep{chen2016empirical, dash2024nudges}. Which seller will end up in the \buybox{} arguably has significant impact on consumer choice, as consumer may simply buy from the seller appearing in the default \buybox{}, rather than comparing competing offers in detail. We explore the implications of Amazon's \buybox{} design in Section \ref{sec:buybox}. 

Second, the seller featured in the Amazon \buybox{} and all other sellers are listed on a separate offer listing page~\citep{dash2024nudges}. This page includes seller performance metrics such as average user ratings, percentage of positive feedback, and the number of ratings. This setup enables consumers to compare the sellers from whom they would like to purchase the product. We investigate the choice architecture of these offer listing pages in Section \ref{sec:offer_listings}. 

Third, Amazon has designed its platform not only for consumers accessing the platform through a computer with a web browser. The platform can also be navigated using voice control: Amazon provides conversational voice search through its smart speakers (Echo devices) powered by the in-house voice assistant Alexa \citep{dash2022alexa}. As consumers accessing Amazon through voice control can only process limited amounts of information, Amazon's choice architecture for voice access presents a drastically reduced version of Amazon's interface to consumers. We explore the implications of some of Amazon's decision in this choice architecture in Section \ref{sec:alexa}. 

Finally, Amazon provides various recommendation systems guiding consumer search and providing additional information about products and seller \citep{sorokina2016amazon, dash2021when}. We explore some of the choice architectures underlying these recommendation systems in Section \ref{sec:recommendation_systems}.

While Amazon uses these and other choice architectures to design its platform, Amazon's role is not restricted to a platform designer. In addition, Amazon sells directly to consumers in most of its marketplaces -- for example, in the United States, France, or Germany -- making it one of the major sellers on its own marketplace. This puts Amazon in direct competition with independent sellers for many products that are sold on Amazon. In some countries -- for example, India -- Amazon is not allowed to sell products on its marketplace directly. As a response, Amazon created joint ventures in India in collaboration with other retailers to sell products on Amazon.in. Cloudtail India and Appario \new{Retail Private Ltd.}, for example, were two joint ventures which emerged as two of the biggest sellers on Amazon India.\footnote{While Cloudtail ceased to offer products on Amazon in 2022, Appario remains in the business of selling products on Amazon. Both companies were sellers on Amazon India while the data presented in this article was collected.} 

Apart from operating the marketplace and being a seller, Amazon also provides fulfillment and logistical services. For example, Amazon stores third-party products in its fulfillment centers and packs and ships them to different customers upon arrival of orders. This service is usually known as Fulfillment By Amazon (FBA) \citep{amazon2020FBA}. For instance, sellers can opt for the full range of services offered in the Fulfillment by Amazon program; they can choose to store their products at their own warehouse and merely rely on Amazon's delivery service; or they can select to store, pack and ship their products either on their own or through other third-party services. By opting to use Amazon's subsidiary services (fulfillment and/or delivery), sellers (with whom Amazon competes on its marketplace) also become Amazon's customers for logistical services.

The intricate relationship between Amazon and sellers on Amazon Marketplace also plays out at the product level. Amazon manufactures its own \textit{private label products} and sells them under different brand names (for example, AmazonBasics, Presto, or Solimo) \citep{dash2021when}. Note that only Amazon sells Amazon's private label products on Amazon. 

Finally, Amazon also maintains an advertising platform. For example, when
a customer looks for a product on Amazon, the platform presents both organic
and sponsored recommendations. While organic recommendations refer to the
results that the algorithmic system produces in the desired context based on
previous customer behavior, sponsored recommendations are sponsored product
advertisements which are the result of second-price auctions in which sellers
pay Amazon so that their product is displayed in a favorable position. Notably,
Amazon also displays its own products among sponsored search results \citep{dash2021when}. 

%% file: 3_Antitrust.tex
As the preceding section has shown, Amazon fulfills multiple roles on its digital platform, acting as a platform designer and operator, seller and product manufacturer, provider of fulfillment and logistical services, and operator of an advertising platform. The way how sellers and customers interact on Amazon's platform is heavily influenced by Amazon's choice architectures, which determines the context in which customers identify products and make purchase decisions.

The complex interactions between multiple players on Amazon are a result of continuous product development and innovation by Amazon over the last three decades. Originally, Amazon emerged as a book seller, entering the business of providing a platform for other sellers only later. Similarly, major online travel agencies such as Booking.com or Expedia over time became vertically integrated with meta-search platforms, enabling these travel agencies to compete with independent competitors on these platforms \citep{cure2022}. 

When a digital platform such as Amazon sells its own products and products from affiliated producers on its online platform, this raises issues of vertical integration. Can a vertically integrated platform that operates both as platform provider and seller sufficiently distinguish between both roles? Or is there a risk that it has incentives to bias search results and recommendations in favor of its own or affiliated products, at the expense of products from competing independent sellers?

Problems of vertical integration are not unique to digital platforms. The practice of supermarkets providing preferential shelf space to their own or affiliated products (see \citealt[202]{tagliavini2023}), and financial advisers earning commissions that might bias their investment recommendations \citep{cookson2021,egan2019}, both raise similar questions. From an economic perspective, vertical integration can have ambiguous effects on social welfare. Vertical integration of digital platforms may increase consumer search costs, discourage product innovation and prompt sellers to leave the market, as preferential treatment by the platform provider for own or affiliated products may make it more difficult for independent sellers to compete (see \citealt{farronato2023}). However, it is not clear that such welfare-decreasing effects will always dominate. By engaging in preferential treatment, a dominant vertically integrated platform provider may have incentives to invest in product quality, product design and innovation. It may also have an incentive to decrease prices and increase output because it can capture a larger fraction of the benefits from such actions (often referred to as the ``internalization of double mark-ups,'' see \citealt[7]{nonhorizontal2008}). Therefore, whether vertically integrated platform providers are beneficial or detrimental to economic welfare is a complex question that is often impossible to answer in the abstract, based solely on theoretical considerations (see, for example, \citealt{corniere2019,hagiu2022,anderson2022}).

These insights from industrial organization research have had an influence on competition law's treatment of vertical integration. With the increasing influence of the Chicago School in antitrust, many jurisdictions have become reluctant to condemn cases of vertical integration as such. Rather, antitrust authorities investigate potential violations of antitrust laws in vertical relationships as part of a rule-of-reason analysis. Under rule-of-reason analysis, antitrust agencies and courts examine both the positive and negative effects of a certain conduct on competition in order to determine whether the conduct violates antitrust laws. For example, U.S. courts have shifted from a \textit{per se} prohibition of resale price maintenance\footnote{Compare Dr. Miles Medical Co. v. John D. Park \& Sons Co., 220 U.S. 373 (1911) (treating a vertical price fixing scheme as per se illegal) with Leegin Creative Leather Prods., Inc. v. PSKS, Inc., 551 U.S. 877 (2007) (holding that all vertical distribution restraints – non-price and price – should be analyzed under the rule of reason).} and territorial restrictions\footnote{Compare United States v. Arnold, Schwinn \& Co., 388 U.S. 365 (1967) (holding that vertical restraints limiting distributors to exclusive sales areas is per se illegal) with Cont’l T.V., Inc. v. GTE Sylvania, Inc., 433 U.S. 36 (1977) (applying a rule of reason analysis to non-price distribution restraints protecting interbrand competition).} to a rule-of-reason analysis that focuses on the economic effects of vertical restraints. Relatedly, self-preferencing by a dominant firm is not a \textit{per se} violation of European competition law (Art. 102 of the Treaty on the Functioning of the European Union, TFEU), but subjects to an effects test \citep[7, 66]{cremer2019}. 

This effects-based approach to assessing vertical integration under antitrust laws has influenced the way authorities have addressed alleged preferential treatment by digital platforms. The  General Court of the European's recent Google Shopping decision provides an example. In this decision, the court reviewed the European Commission's finding that Google had promoted its own comparison shopping service on its general search engine at the expense of rival services (see General Court, Case T-612/17, Nov. 10, 2021, ECLI:EU:T:2021:763). In its decision, the court pointed out that not all foreclosure is detrimental to competition, and that the mere extension of a dominant position to a neighbouring market is not necessarily anti-competitive (General Court, Case T-612/17, Nov. 10, 2021, ECLI:EU:T:2021:763, para. 157, 162). While the court ultimately confirmed the European Commission's decision, the case demonstrates that whether self-preferencing violates European competition law can only be decided on a case-by-case basis. This analysis involves various different competition doctrines such as duty to deal, tying, margin squeeze, general abuse, or discrimination (see \citealt{colangelo2023}).

Antitrust's rule-of-reason analysis indicates an important feature of this area of the law: antitrust law traditionally focuses on ex-post regulation. Antitrust agencies and courts enforce rules, but they typically do this in a reactive manner. They investigate firm behavior and issue decisions to hold firms accountable for past non-compliance and to ensure compliance in the future. 

Over the recent years, digital platforms have become subject to increased scrutiny with regard to their vertical integration. This includes their role as a platform provider, product manufacturer and seller, logistical service provider, and advertising platform operator. There has been a growing concern that antitrust law's focus on ex-post intervention may not be sufficient to ensure viable competition in platform markets. In particular, network effects and switching costs might make it effectively impossible for sellers and customers on platform to switch to a competing platform, thereby preventing competition in the market for digital platforms: it may just not be profitable for a competitor to Amazon and other digital platforms to create a competing marketplace. 

In this environment, calls to introduce an ex-ante regulatory regime gained prominence. In such a regime, rules and standards are established by the regulator before issues arise, and firms have to obey them when they become active in the market. When a particular conduct is prohibited, the regulator does not allow agencies or courts to take countervailing evidence into account, as is typical in rule-of-reason analysis. Once the regulator has determined that certain firm conduct is harmful, for example, this conduct is prohibited, even if the firm could show that the conduct also had beneficial effects on economic welfare.

Along these lines, the European Union recently introduced a regulatory system for digital markets that is based on an ex-ante approach where efficiency defenses raised by firms play a very limited role. With the Digital Markets Act (DMA), which came into full force in 2024, the European Union aims to make key digital markets fairer and more contestable by preventing large companies from abusing their market power and lowering barriers to the entry and expansion of smaller players. The DMA targets large digital platforms -- so-called ``gatekeepers'' -- such as Alphabet, Amazon, Apple, Meta, or Microsoft and imposes various obligations on them. 

Article 6(5) of the DMA, for example, includes an explicit prohibition according to which gatekeepers ``shall not treat more favourably, in ranking and related indexing and crawling, services and products offered by the gatekeeper itself than similar services or products of a third party. The gatekeeper shall apply transparent, fair and non-discriminatory conditions to such ranking.'' Importantly, this provision does not allow gatekeepers to provide justifications showing that any economic disadvantages created by self-preferencing are outweighed by pro-competitive effects. Efficiency defenses play no role in such ex-ante regulatory approach. 

Article 6(2) DMA prohibits digital platforms from using, in competition with other sellers, any non-public data that is generated or provided by those sellers. This prohibition is intended to countervail the ability of vertically integrated digital platforms to observe sales data from independent sellers with whom the platforms compete. The DMA also includes, among other provisions, prohibitions on combining data from different services belonging to the same company, protections for advertisers and publishers using the gatekeeper against monopolization trends, and provisions for ensuring interoperability and data access for business and individual users of platforms. If gatekeepers fail to abide by these obligations, the European Commission may impose fines of up to 10 \% of the gatekeepers' annual worldwide turnover (Art. 30(1) DMA). The Digital Market Act empowers the Commission to monitor the effective compliance of digital platforms with the obligations of the DMA, potentially using independent external experts, auditors and member state authorities (Art. 26 DMA). Article 21 DMA entitles the European Commission to require access to data and algorithms of digital platforms. 

Going beyond the Digital Markets Act, the European Union also enacted the Digital Services Act (DSA), which came into effect in 2024 as well. This regulation provides a comprehensive framework regarding illegal content, transparent advertising, and disinformation of online platform providers. Article 26 of the DSA, for example, requires online platforms to be transparent about the nature and origin of advertisements. This obligation also applies to ``sponsored links'' displayed by digital platforms. Articles 25 and 27 of the DSA include rules about online interface design and recommender system transparency. The European Commission may impose fines of up to 6 \% of the annual worldwide turnover of very large online providers (Art. 74(1) DSA), which includes Amazon, Alibaba, Apple, Google Shopping, and Zalando, among other providers.\footnote{For a list of very large online platforms as designated by the European Commission, see \url{https://digital-strategy.ec.europa.eu/en/policies/list-designated-vlops-and-vloses}.\new{accessed April 02, 2024}}

%% file: 4_Preferential_Treatment.tex
The effectiveness of regulatory action against platforms such as Amazon will always depend on the extent to which their behavior can be observed. While public discussions about these platforms often rely on individual cases or on anecdotes, algorithmic enforcement tool might open up a way to investigate potential self-preferencing behavior in a systematic, quantifiable manner. Doing so requires tools to observe Amazon's choice architectures and their effects on costumer behavior in an automated, scalable manner.

In the following section, we summarize a series of empirical investigations
into various of Amazon's choice architectures undertaken by several co-authors
of this article \citep{dash2021when, dash2022alexa, dash2024nudges}. The purpose of this section is to
describe their main findings and distill important insights about the promises
and challenges of algorithmic enforcement of platforms.

In the following section, we present the results of several empirical investigations into various of Amazon's choice architectures. The underlying analysis was conducted by several co-authors of this article. The purpose of this section is to summarize their findings and distill important insights about the promises and challenges of algorithmic enforcement of platforms. 

\subsection{Amazon Buy Box}\label{sec:buybox}
\input{BuyBox/BuyBox.tex}

\subsection{Offer Listing Pages}\label{sec:offer_listings}
\input{OLP/Offer-Listing-Page.tex}

\subsection{Alexa Search}\label{sec:alexa}
\input{Alexa/Alexa.tex}

\subsection{Recommendation Systems}\label{sec:recommendation_systems}
\input{IR/IR.tex}

%% file: BuyBox/BuyBox.tex
\begin{figure}[tb]
	\centering
        \caption{\textbf{Example of a \buybox{} on Amazon}}
            \label{fig:buybox}
		\includegraphics[width=0.3\textwidth, height=4.5cm]{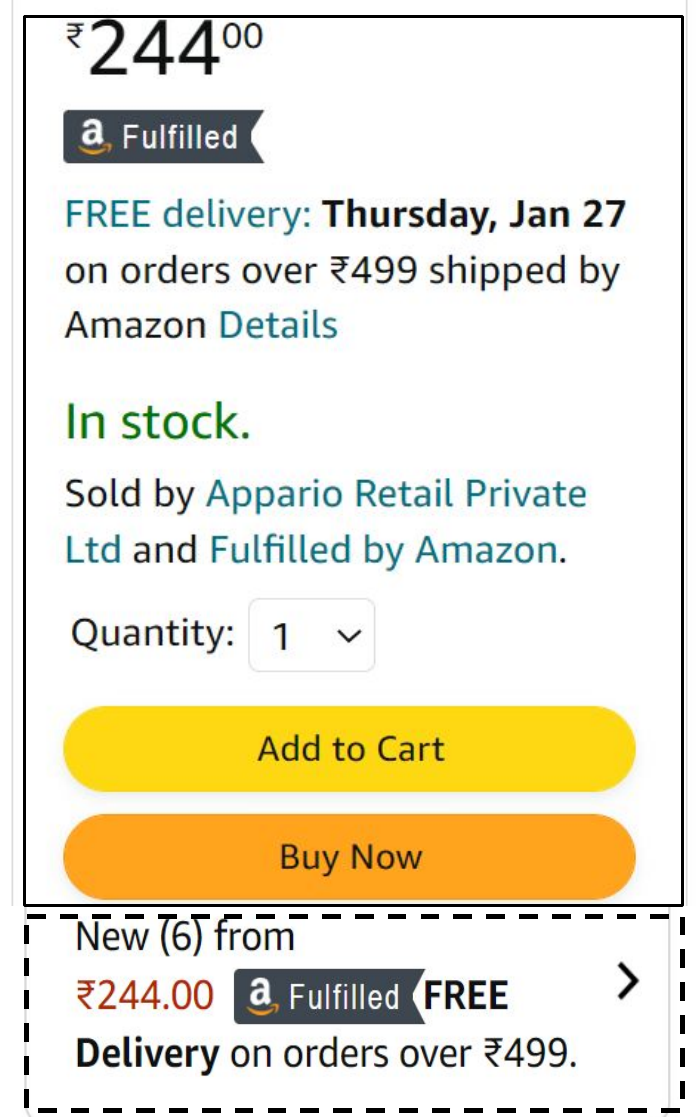}
		
	
	\vspace*{-5mm}
\end{figure}

We start our exploration of Amazon's market practices by investigating how the service decides which offer to feature in its  Buy Box\footnote{Summarized from \citet{dash2024nudges}.} (now called ``Featured Offer''). A typical Amazon product page has a rectangular box on the top-right corner, right next to the title of the product. This box is popularly known as the \buybox{}. Figure~\ref{fig:buybox} shows an example of such a \buybox{}. The box contains the price of the product, a delivery estimate, and two buttons that read: 
\new{``Add to Cart'' and ``Buy Now''}. 
The box also lists the name of a seller. In the \buybox{} depicted in Figure~\ref{fig:buybox}, the seller is 
\new{Appario Retail Private Ltd}.

In principle, there are various ways through which customers can buy products on Amazon. However, given the prime positioning of the \buybox{} (and the prominently displayed buttons inside the box), it accounts for more than 80\% sales on Amazon \citep{Lanxner2021Amazon}. 

Given the importance of the \buybox{} for transactions on Amazon's platform, Amazon exerts considerable power over customers' purchase decisions through its algorithm that selects the seller that appears in the \buybox{}. This decision is trivial if there is only one seller that offers a given product. However, in practice, multiple sellers can and often do offer the same product. In this scenario, Amazon selects the seller to be featured in the \buybox{} through a proprietary algorithm known as the \textit{Featured Offer Selection} algorithm, also commonly known as the \buybox{} algorithm. The precise structure of this algorithm is unknown, making it a subject of speculation (and concern) among sellers on Amazon \citep{Kaneshiro2020Keeping}. Amazon presents some clues and recommendations on its seller interface regarding what factors its algorithms take into
account in determining the winner of the \buybox{}. These factors include, for example, competitive pricing and seller performance metrics. 

Amazon's \buybox{} algorithm has been at the center of multiple antitrust investigations. A main concern is that Amazon might design the algorithm in a way that amounts to self-preferencing. To understand what that means, remember that Amazon directly competes with other business users for selling products on its marketplace. As discussed in Section~\ref{sec:Amazon}, Amazon also provides paid logistical services to sellers who opt for it, raising the question whether sellers that purchase Amazon's services receive preferential treatment.

We start our investigation by analyzing how frequently Amazon (or its special merchants in India, see Section \ref{sec:Amazon}) win the \buybox{}. To answer this question, we collected data for over 70,000 buy box competitions in Amazon's Indian, American, German and French marketplaces. Specifically, we recorded the top-100 popular queries on Amazon and searched for them on each of the marketplaces. We then recorded, for each marketplace, all the products that appeared on the first page of the search result page. For all products, we proceeded to the product page and checked whether multiple sellers offered the corresponding product. If the answer this question was yes, we collected information about the sellers and the winner of the \buybox{}. Since \buybox{} winners can potentially change between different queries, we collected the data for a span of two weeks. 

During our period of observations and across all the countries analyzed, we observe that Amazon won upwards of 80\% of the \buybox{} competitions where it was involved as a seller. At the same time, only 11\%, 13.22\%, 17.74\%, and 26.23\% of the sellers competing with Amazon got to win at least one Buy Box competition throughout our data collection period in the Indian, US, German and French marketplaces, respectively. Interested readers can find more details about the data collection and analyses in \citet{dash2024nudges}.

Of course, the fact that Amazon won most \buybox{} competitions on its own does not prove self-preferencing behavior on the part of the platform. After all, it appears possible that Amazon's own offers are preferred by its customers. If that was the case, the \buybox{} algorithm would simply give customers what they want, not nudge them towards purchase decisions that benefit Amazon at the expense of its customers and competing sellers. 

To investigate this question, we first compare the price offered by Amazon and rival sellers in cases in which Amazon won the \buybox{}. We find considerable numbers of instances in which Amazon won the \buybox{} despite offering the product at a price that is higher than that offered by at least one rival seller: In the US and in India, 25\% of the \buybox{}es won by Amazon and its special merchants are cases where Amazon did not offer the corresponding product at the lowest price. In the European marketplaces, these numbers were slightly lower (12.45\% for France and 13.91\% for Germany).

However, information on offer prices cannot exclude the possibility that the results of the \buybox{} algorithm, despite its apparent preference for Amazon over other sellers, produces results that map to consumer preferences. If the outcomes of \buybox{} competitions usually match with consumer preferences, Amazon's choice architecture might be designed in a welfare-maximizing way. 

Determining consumer preferences in digital markets in a reliable and reproducible manner is a challenging task. The data available to us from our empirical investigation of Amazon is insufficient to answer the question whether the \buybox{} algorithm's results correspond to consumer preferences. We therefore turn to a different empirical method that allows us to gain insights into consumers preferences: Using the Prolific crowdsourcing platform, we conducted a survey among 200 respondents (50 from each of the four countries). 

In the survey, we showed survey participants from each country the four best offers for a product and elicited their preference regarding the different sellers. We started by randomly selecting, for each country, 15 products from among the products for which Amazon, even though it did not offer the lowest price, won the \buybox{} in the first iteration of the data collection. For each product, we showed our survey participants the four offers with the lowest price and delivery charges. For each of the four selected offers, we displayed the name of the seller, the price offered (including delivery charges, if any), the average user rating of the seller, the percentage of positive feedback of the seller, the  number of ratings received, and delivery options.\footnote{We obtain this information from the offer listing page for that product, which will be discussed in greater length in Section~\ref{sec:offer_listings}. For the present context, it suffices to say that Amazon displays all competing sellers on a separate page, together with information about the seller, product price and delivery charges.} We then asked participants the following question: ``Suppose you are willing to buy a `<\textit{product name}>' on Amazon and these are the offers from different sellers for the same product. Which one would you prefer to buy?'' 

\begin{figure}[tb]
	\centering

        \caption{\textbf{Amazon Buy Box and Survey Participants}}
        
		\includegraphics[width=0.5\textwidth, height=4.5cm]{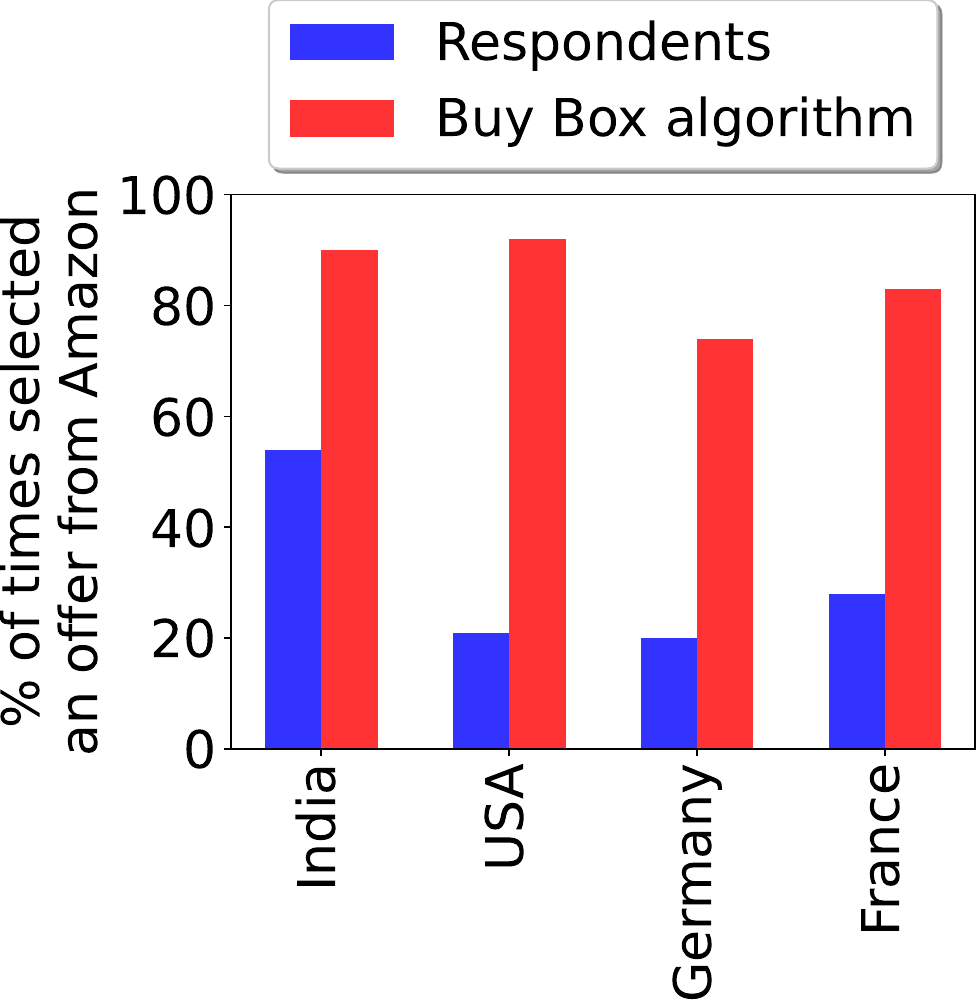}
		
	\scriptsize{\textit{Note:} This figure shows how often survey participants and Amazon's \buybox{} algorithm selected an offer from Amazon for the surveyed products. We observe a significant gap in survey participants' preference as opposed to that of the algorithm's.}
    \label{fig:buyboxSurvey}
    \vspace*{-5mm}
\end{figure}

Out of a total of 3,000 evaluations, participants chose the offer from Amazon as their first preference in only 31\%. This preference varied across countries (as shown in Figure~\ref{fig:buyboxSurvey}), with 54\% in India, 21\% in the USA, 20\% in Germany and 28\% in France. In contrast, if we consider the winner of the Buy Box for the 60 products surveyed at different time points that we observed during our data collection, more than 80\% of the times the Buy Box had been won by the Amazon (90\% in India, 92\% in the USA, 74\% in Germany, and 83\% in France). This suggests a considerable mismatch between the frequency at which the \buybox{} algorithm selected an offer from Amazon and the frequency at which survey participants selected the same~\citep{dash2024nudges}. 

This result raises doubts about whether consumer preferences align with seller choices determined by Amazon's \buybox{} competitions. By contrast, it might suggest a potential pattern of self-preferencing in the \buybox{} algorithm in scenarios in which other sellers offer the product at a lower price than Amazon. In these scenarios, the \buybox{} algorithm might nudge consumers towards Amazon as a seller.

%% file: OLP/Offer-Listing-Page.tex
\begin{figure}[b]
	\centering
        \caption{\textbf{Offer Listing Page on Amazon: Example}}
		\includegraphics[width=0.6\textwidth, height=4.5cm]{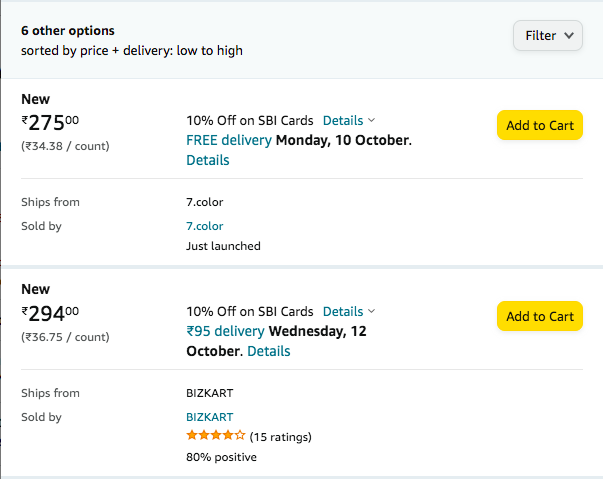}
		
	
    \label{fig:OLP}
	\vspace*{-5mm}
\end{figure}

We continue our exploration of Amazon's market practices by investigating its offer listing page, another choice architecture in the Amazon ecosystem\footnote{Summarized from \citet{dash2024nudges}.}. As described in the previous section, product pages on Amazon include a \buybox{} that allows for quick product purchases and for which Amazon, if necessary, selects a seller from among the competing sellers offering a product. However, consumers are not bound to use the \buybox{}. Amazon offers consumers a separate page which lists all the available offers for a product. Figure~\ref{fig:OLP} displays an example of such an offer listing page. To get to the offer listing page, consumers can click a link that appears in a second rectangular box displayed below the \buybox{}, see Figure~\ref{fig:buybox}. The offer listing page shows the offers from all the competing sellers ordered by price and delivery charges. It also shows some seller performance metrics including its average user rating, the percentage of positive feedback, and the number of ratings.

In our empirical investigation, we focus on two specific characteristics related to how Amazon displays seller performance metrics~\citep{dash2024nudges}. First, while some metrics provide an indication of the (average) quality of service offered by a seller, the number of ratings (\#Ratings) reflects the total feedback ratings received by a seller throughout their tenure on Amazon. The inclusion of such metrics could potentially sway customers towards larger, more established sellers. It is important to recognize that inferring anti-competitive intent solely from this design choice is challenging, as consumers may naturally gravitate towards proven sellers for a sense of reliability and trustworthiness.

\begin{figure}[b]
	\centering
        \caption{\textbf{Review Striking Through: Example}}
		\includegraphics[width= 0.6\textwidth, height=4.5cm]{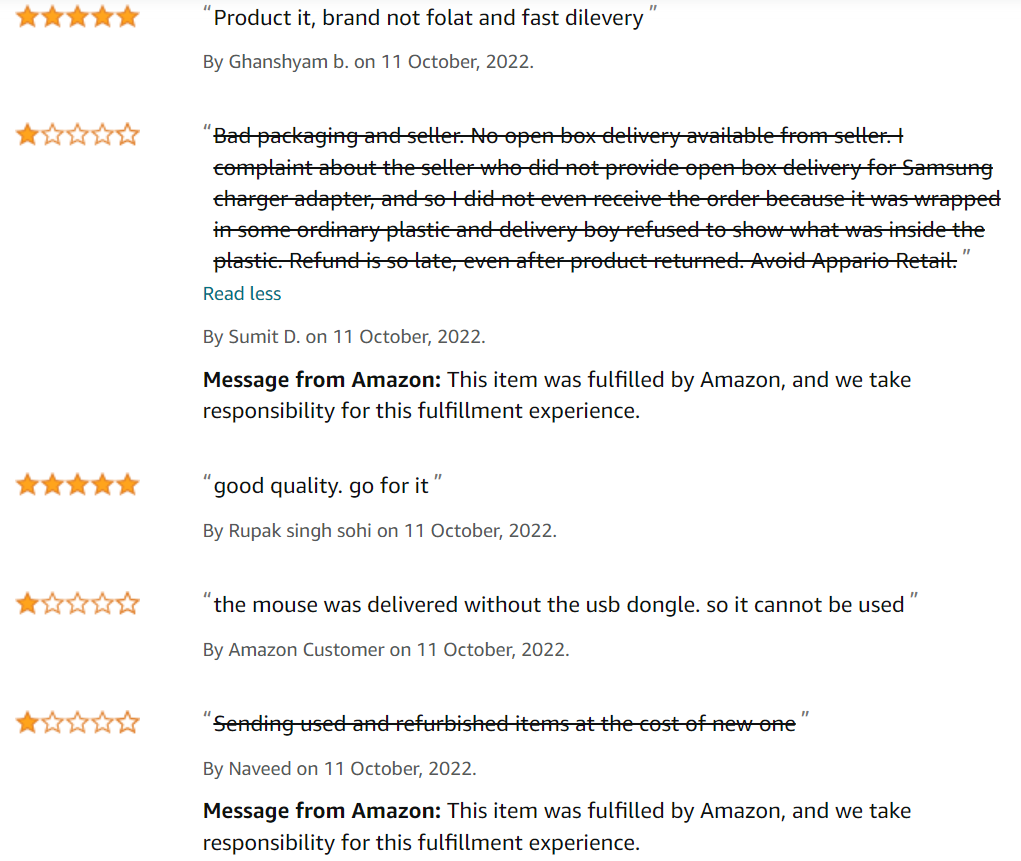}
        \label{Fig: SellerProfilePage}
        \newline
	\scriptsize{\textit{Note:} This figure provides an example of review strike throughs: the second and the last reviews have been struck through. A message from Amazon is mentioned taking responsibility of the issue.}

	\vspace{-6mm}
\end{figure}

Second, we turn to a specific feature of Amazon's approach to displaying customer ratings on the offer listings page that might be less innocuous, specifically because it might offer advantages to sellers who use its subsidiary fulfillment and/or delivery services (see Section \ref{sec:Amazon}). For these sellers, under certain circumstances, Amazon reserves the right to ``strike through'' negative feedback by consumers. For example, Amazon may strike through a seller feedback if ``the entire comment relates explicitly to delivery experience for an order fulfilled by Amazon '' or ``the entire comment is related to a delayed or undelivered order, which you shipped on time by using Buy Shipping''~\cite{SellerCentral2022Can}. In both cases, in addition to the strike through, Amazon puts up a message taking responsibility for the inconvenience faced by the customer. Figure~\ref{Fig: SellerProfilePage} shows examples of struck-through reviews. Amazon states in its Fulfilled-by-Amazon and Amazon Shipment advertisements that negative reviews which have been struck through will not impact sellers' performance metrics \citep{SellerCentral2022BuyShip, SellerCentral2022Buyer}. Importantly, this policy is unavailable for independent sellers that do not use Amazon's fulfillment or delivery services. Hence, for the same customer inconvenience, sellers might be treated differently by Amazon depending on whether they purchase services from Amazon or not. 

To investigate the impact of the strike-through policy on Amazon seller performance metrics, we collected feedback data for the top-1000 active sellers in Amazon's Indian, US, German and French marketplaces. Our dataset consists of a total of 4 million seller feedback reviews for these 4,000 sellers. The data collected include the feedback text, associated ratings and whether it was struck through or not by Amazon. More details on the data collection can be found in \cite{dash2024nudges}. 

We categorized the sellers in our dataset into three main groups: (1)~Amazon Fulfilled, (2)~Amazon Shipped, and (3)~Merchant Fulfilled. The determination of seller membership in the first two groups was based on the text displayed when Amazon strikes through reviews. Sellers falling into the third group were identified if none of their reviews in the dataset were struck through. Subsequently, we conducted a manual inspection of their product pages to ascertain how they fulfill their orders.

We then assessed how often Amazon struck through seller feedback pertaining to members of the first group, and how the striking through of feedback affected the relative ratings of Amazon Fulfilled sellers as opposed to Merchant Fulfilled sellers. Notably, more than half of the reviews in which Amazon Fulfilled sellers received only three stars or less was struck through in every country we investigated. For Amazon Shipped sellers, the numbers hover between 22\% and 4\% dependent on the country. Reviews for Merchant Fulfilled Sellers were never struck through. 

The exclusion of struck-through feedback markedly changes how different types of sellers perform in ratings. For example, when excluding struck-through reviews, the overall ratings are best for Amazon Fulfilled sellers. When struck-through reviews are included in the evaluation, Amazon Fulfilled appear to have lower rankings than Merchant fulfilled sellers in India, the US, and Germany.

Consistent with our previous investigations, these findings suggest that Amazon's choice architecture favors sellers affiliated with Amazon. Different from the inclusion of the number of ratings as a performance metric, it seems less likely that the decision to exclude certain disfavorable ratings for Amazon-affiliated sellers from the calculation of seller performance is aligned with consumer preferences. At the very least, it results in the display of ratings that are not comparable across different types of sellers, thereby reducing their informativeness to consumers. At worst, it could distort consumers' purchase decisions.

But do these design decisions have a real impact on consumer's purchase behavior? To investigate this question further, we again conducted a survey, this time with participants in India. In the survey, we exposed participants to several counterfactual settings. We showed them (a)~the Amazon-reported seller metrics with number of ratings (\# Ratings), (b)~the Amazon reported-metrics seller without \# Ratings, (c)~the rectified metrics (including the struck-through feedback) with \# Ratings and, finally, (d)~rectified metrics without \# Ratings. We recorded the percentage of times respondents selected the offers from Amazon special merchants or Amazon fulfilled sellers in each of the settings. The results are listed in Table~\ref{Tab: ImpactofFeatures}. 

The results of this survey suggest that when participants saw the rectified seller metrics, their preference toward sellers using Amazon's logistical services reduced significantly in all cases. Similarly, when the number of ratings were withheld, the participant's preference toward Amazon special merchants reduced by nearly 40\%. 

In summary, the observations in this section indicate that various aspects of Amazon's offer listings page, including its strike-through policy for negative reviews, have a divergent impact on different types of sellers, with those affiliated with Amazon receiving preferential treatment. Additionally, our findings suggest that these design choices can significantly influence consumers' purchase decisions.

\begin{table}[!t]
	\noindent
    \footnotesize
	\centering
        \caption{\textbf{Preferences by Survey Participants}}
	\begin{tabular}{ |p{5.5 cm}|p{2.75 cm}|p{2.75 cm}|}
		\hline
		{\bf \new{Percentage} of 1st Preference Votes} & {\bf Amazon Metrics} & {\bf Rectified Metrics}  \\
		\hline \hline
        \multicolumn{3}{|c|}{\textbf{Amazon Special Merchants}}\\
        \hline
		  With \#Ratings & 54.26\%  & 28.53\% \\ 
		\hline
		Without \#Ratings & 14.53\% & 6.93\% \\ 
		\hline \hline
        \multicolumn{3}{|c|}{\textbf{Amazon Fulfilled}}\\
        \hline
		With \#Ratings & 81.20\%  & 64.53\% \\ 
		\hline
		Without \#Ratings & 72.13\% & 54.6\% \\ 
		\hline
	\end{tabular}	
	\scriptsize{\textit{Note:} This table shows the percentage of first preference votes aggregated by Amazon Special Merchants and Amazon Fulfilled sellers in the different survey settings. We observe that the percentages drop when we show rectified metrics as opposed to metrics shown by Amazon or when we omit the \#Ratings feature.}
	\label{Tab: ImpactofFeatures}
	\vspace*{1mm}
\end{table}

%% file: Alexa/Alexa.tex
Next, we shift our focus to product selections facilitated by Amazon's virtual assistant, Alexa~\footnote{Summarized from \citet{dash2022alexa}.}. While many consumers traditionally purchase products from Amazon through its website, there exists an alternative avenue: Alexa enables consumers to make purchases through voice commands, interacting with smart speakers.

\begin{figure}[hbt!]
	\centering
        \caption{\textbf{Amazon Alexa}}
        \label{fig:alexa}
		\subfigure[Search through Alexa]{\includegraphics[width=0.3\textwidth, height=4.5cm]{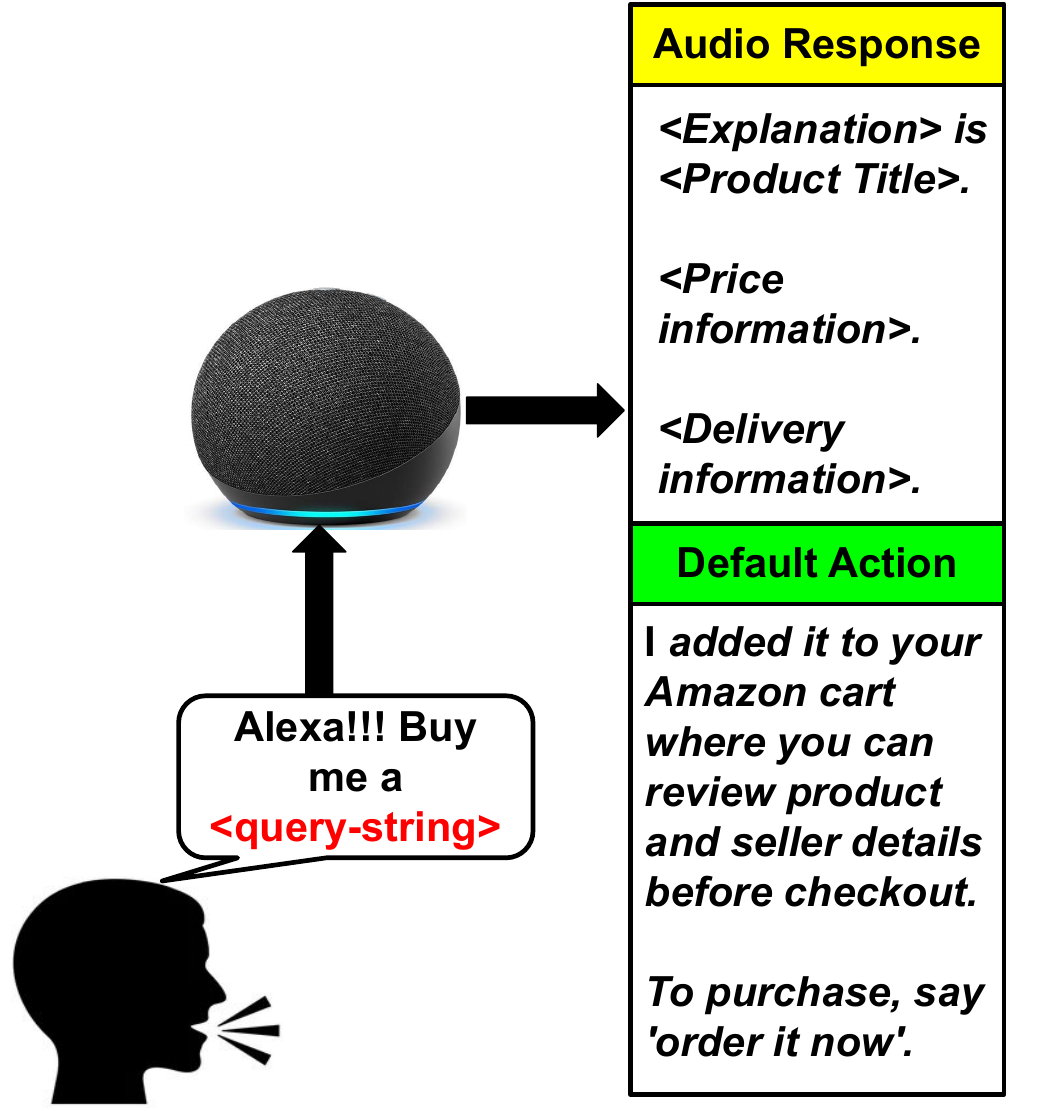}}
        \subfigure[Data Collection Pipeline]{\includegraphics[width=0.5\textwidth, height=4.5cm]{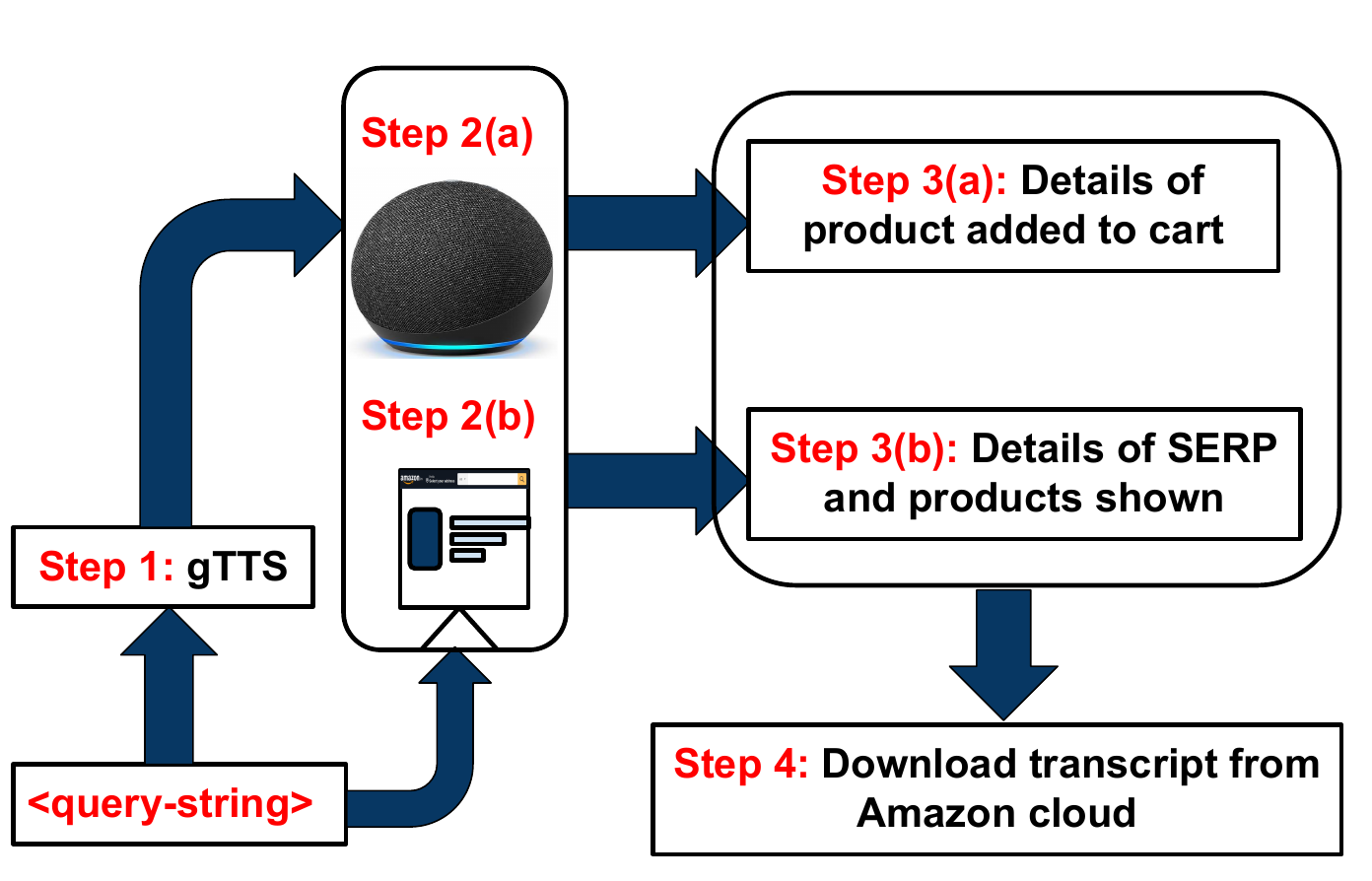}}
		
	\scriptsize{Panel (a) shows the response to a product search through Alexa: In response to a purchase query, (i)~the virtual assistant spells out an audio response explaining a product information, and (ii)~adds the product to cart for further exploration. Panel (b) shows the data collection pipeline: a query string gets converted to an audio signal by Google Text-To-Speech. The audio signal and the textual query are provided to Alexa and desktop search respectively. Finally, the product page details of the retrieved products and the transcripts of the conversation are collected.}
    \vspace*{-5mm}
\end{figure}

Figure~\ref{fig:alexa}~(a) graphically depicts a query and a typical response from Amazon Alexa. When a consumer asks a smart speaker powered by the Alexa virtual assistant to purchase a product online, the typical responses has two main parts: (1)~an \textit{audio response describing a chosen product with a brief explanation}, and 
(2)~a \textit{status quo or default action}. As part of its product description, Alexa spells out relevant product details, including its title, price and delivery information. Often, the response also contains a brief explanation of why the virtual assistant has chosen the corresponding product. Explanations provided by the virtual assistant include that the product is ``Amazon's Choice'' or ``a top result.'' Alexa's default action is to add the product to the customer's cart for further review or purchase. In addition, the virtual assistant asks the consumer whether she wants to make the purchase immediately. 
 
\vspace*{5mm}

Compared to consumers interacting with Amazon on a website, consumers enjoy a significantly smaller action space when interacting with Amazon through a virtual assistant such as Alexa. Rather than presenting a list of search results, Alexa spells out the details of a single product and adds it to the consumer's cart. By reducing consumer choice, Amazon's default choices can have a particularly strong influence on consumer decisions, even stronger than the selection of a seller in the \buybox{}.

Similar to our previous investigations, we start our investigation by collecting information on the products that Alexa selects in response to searches and how these products compare with the search results of a similarly worded search on Amazon's webpage. Overall, we collected search results for 1,000 different queries from both Amazon Alexa and from Amazon's website. We also collected the metadata (including product title, ratings, price, and seller information) for the products added to the cart by Alexa and for products appearing among the search results on Amazon's website. The exact data collection framework is shown in Figure~\ref{fig:alexa}~(b).

We then conducted surveys with 100 participants to elicit consumers' preferences between the product added to cart by Alexa and the top search result (appearing in the first position) from a desktop search. We selected the top search result for comparison because Amazon identified it as the most relevant product for the query at the time for the customer. We showed our survey participants the titles of the two products, their prices, their average user ratings and the number of ratings received by each of the two products \new{--} information we obtained from the Amazon results page during our data collection. We then asked participants the following question: \textit{Suppose you are looking for ``query string''. Which of the following would you prefer to buy?} 

All 100 participants evaluated ten such queries. Out of the 1000 evaluations we received (100 respondents $\times$ 10 queries), 732 survey participants (73.2\%) chose the top desktop search result over the product added to cart by Alexa for the corresponding queries. Analogous to what we find in our investigations of Amazon's \buybox{} algorithm, this might offer suggestive evidence that a majority of participants prefer product offerings that differ from Alexa's choices. 

Overall, we found that Alexa's product selection mechanism often selected products that were different from the top search results on Amazon's webpage, and that, in a majority of cases, consumers would have preferred the top search result over Alexa's product choice. Depending on the identity of the products' sellers, these choices might constitute another example of self-preferencing or a preferential treatment of Amazon-affiliated sellers.

%% file: IR/IR.tex
This subsection turns to yet another choice architecture by which Amazon may engage in self-preferencing: recommendations on Amazon~\footnote{Summarized from \citet{dash2021when}.}.

\begin{figure}[hbt!]
	\centering
        \caption{\textbf{Related Item Recommendations on Amazon}}
        \label{Fig: SearchRIR}
                \includegraphics[width= 0.7\textwidth]{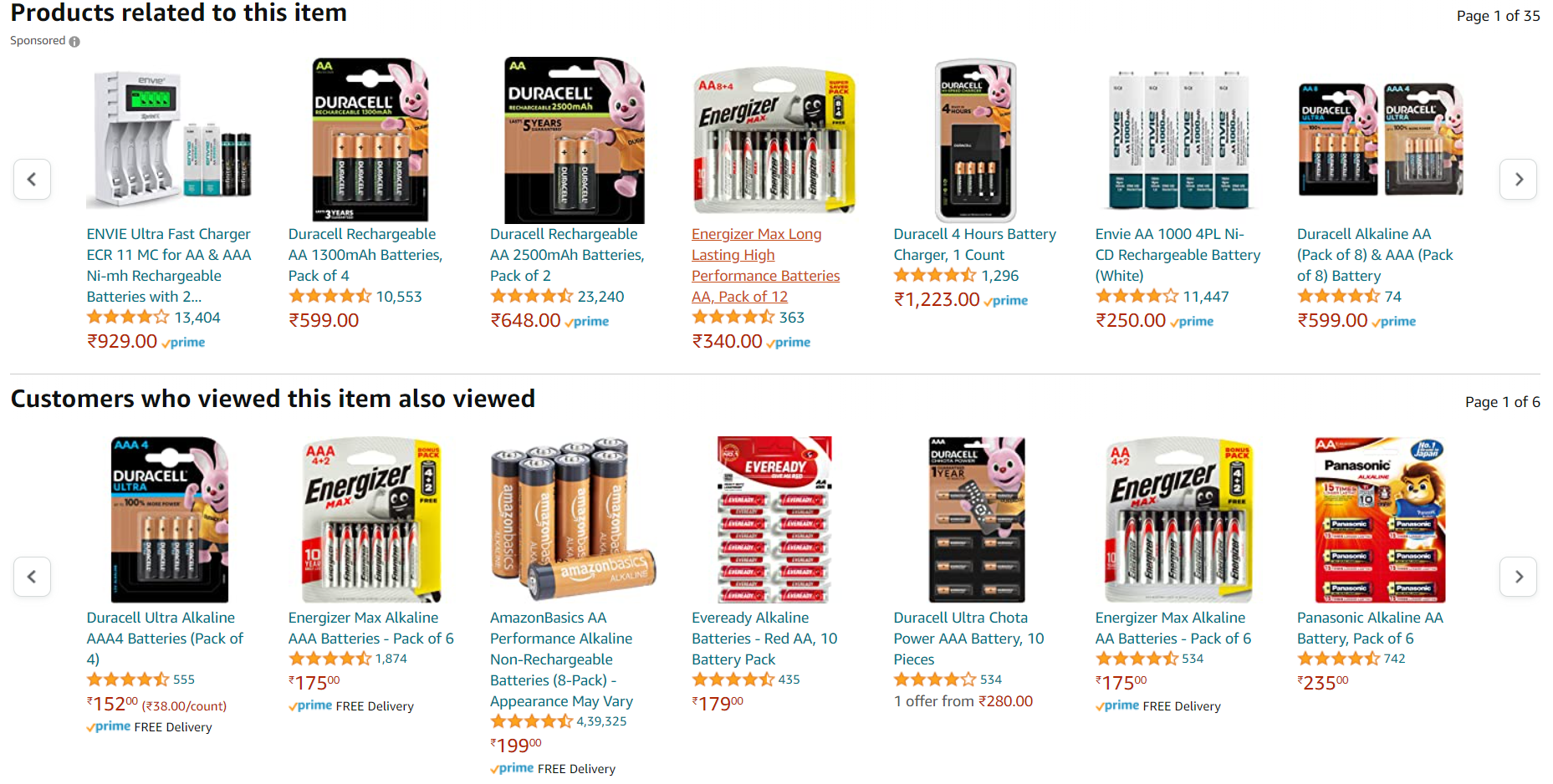}
        \newline
        \scriptsize{\textit{Note:} This figure provides an example of  related-item recommendations on Amazon.
        }
\end{figure}

As an e-commerce marketplace, Amazon provides basic product recommendation functions on its platform (see Section \ref{sec:Amazon}). It seems reasonable to assume that the results of such recommendations will often be a relevant determinants for consumers' purchase decisions. These functions may offer Amazon yet another opportunity to ``push'' its products at the expense of products manufactured and/or sold by others.

Figure~\ref{Fig: SearchRIR} shows a screenshot of related item recommendations. 
These related item recommendations appear towards the bottom of almost all Amazon product pages. Naturally, the number of products Amazon can display in this context is limited, and a crucial question is how Amazon's algorithms decide which products a customer can see.

Amazon displays both (a)~organic results, which appear to be determined on the basis of prior customer behavior, and (b)~sponsored results, with the items displayed determined through second-price auctions in which sellers can participate.\footnote{For the example of related item recommendations, see the lower carousel of recommendations in \new{Figure~\ref{Fig: SearchRIR}}. The title of the carousel reads ``customer who viewed this item also viewed.'' As the title suggests, the items in the carousel are essentially similar to the product on whose product pages they appear based on how customers viewed the items in prior exploration sessions. As has been described elsewhere, these results are curated from item-item collaborative filtering processes \citep{linden2003amazon, smith2017two}).} When Amazon displays sponsored results, it marks those with a tag that reads, on its English language platform, ``sponsored.''

In recent years, Amazon appears to have given more room in its related item recommendations to sponsored results. For example, consider the upper carousel in \new{Figure~\ref{Fig: SearchRIR}}: although the title reads ``products related to this item,'' there is a small text in gray color toward the top left of the carousel that reads ``sponsored.'' This essentially means that the entire carousel does not consist of organic recommendations, but that these are all sponsored product advertisements. 

Of course, Amazon is not the only platform that shows sponsored results alongside organic results. Also, given that sponsored results are selected through a different mechanism than organic results, it is to be expected that the inclusion of sponsored results among related item recommendations will result in fewer organic results shown to the customer, which might be perceived as depriving the customer of valuable information. However, given Amazon's vertical integration, we are specifically interested in whether Amazon uses the sponsored results architecture to promote its private label products.

To analyze this question, we investigate the products shown on Amazon's websites as related product recommendations for batteries. We collected more than 100,000 related item recommendations from the product pages of 5,352 products in the battery category of Amazon using a breadth first search crawler. Out of these 5,352 products, only 17 products (0.3\%) were Amazon private label products. 

\begin{figure}[hbt!]
	\centering
        \subfigure[All Products]{\includegraphics[width= 0.48\textwidth
        ]{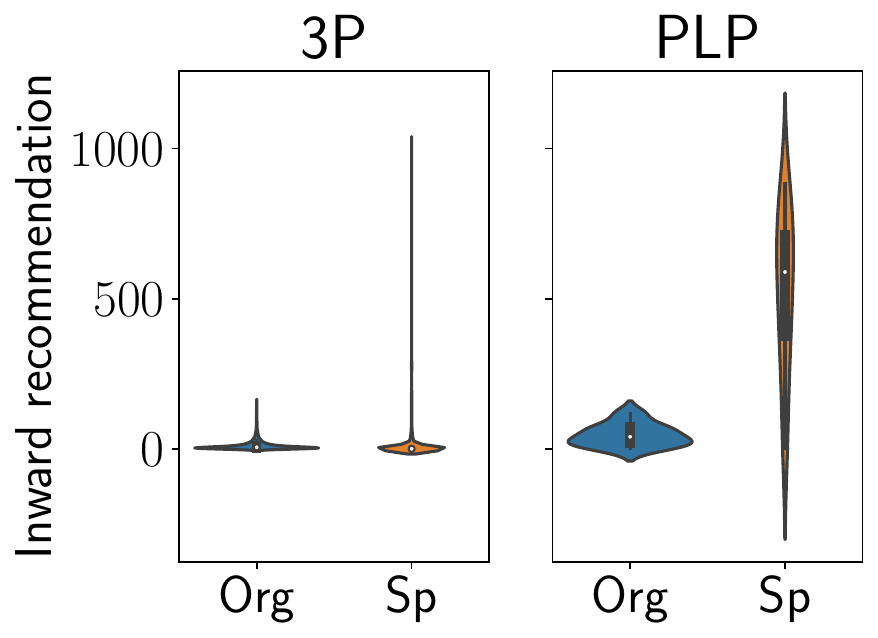}}
        \subfigure[Sponsored Products Only]{\includegraphics[width= 0.48\textwidth
        ]{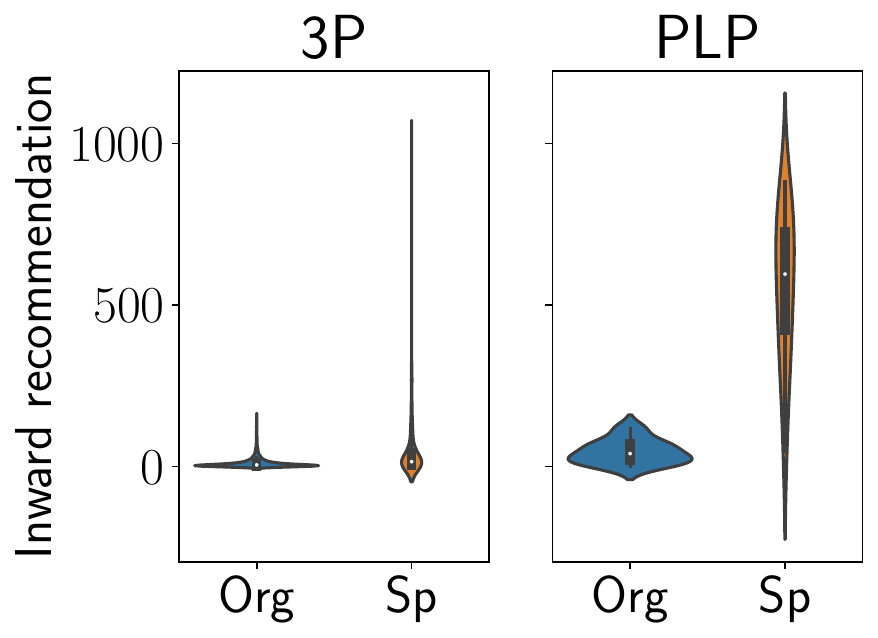}}
        \caption{{\bf \textit{Note:} This figure shows the distribution of inward recommendations toward Amazon private label products (PLP) and third party (3P) products. Panel (a)~depicts distributions for all products, panel (b)~for products that appear at least once as sponsored results.
        }
	}
	\label{Fig: SearchRIRResults}
\end{figure}

We focus on demonstrating a disparity in recommendations that Amazon private label products receive as opposed to third-party products in sponsored vs. organic recommendations. Figure~\ref{Fig: SearchRIRResults} demonstrates that Amazon private label products on average received a much higher number of recommendations among sponsored results (520) as compared with organic product recommendations (46). In other words, Amazon private-label batteries were recommended almost 11 times (520 sponsored recommendations compared to 46 organic recommendations on average) as frequently in the context of sponsored results than in the context of organic results. Third-party products do not experience a similar increase in appearances as recommended products among sponsored recommendations. Such products were recommended, on average, 11 times in organic results. This number does not change much for sponsored results. In fact, a large number (4,357 out of 5,335) of third party products do not receive any sponsored recommendations, thus dragging the mean number of recommendations to 9. By contrast, almost all private label products (all but one) are featured among sponsored recommendations.

Panel (b) shows that, even if one limits the analysis to products that are featured in the sponsored recommendations, the increase in recommendations is bigger for private label products than it is for third party products. The average number of recommendations for the former is 46 among organic recommendations and 553 among sponsored recommendations. For the latter, the number increases from 13 for organic recommendations to 46 among recommendations.

Another way to look at these numbers is to compare how often private-label products appear as recommended products in the context of organic recommendations vs. sponsored results. Amazon's own organic recommendation engine recommended Amazon private-label products from approximately 15\% of recommendations shown to consumers. Simultaneously, they received recommendations from almost 50\% of the product base in the sponsored recommendations. 

Accordingly, it appears that Amazon, by virtue of the sponsored recommendations,  made its private-label batteries much more accessible to consumers looking at other products' web pages. To the extent that the organic recommendation engine captures consumer preferences regarding what other product they consider interesting alternatives to the product they are looking at, this could be viewed as yet another choice architecture that implements a type of self-preferencing.

%% file: 5_Discussion.tex
In this article, we explore unique design features of digital platforms, focusing on those that depend heavily on algorithms and choice architectures. We address regulatory challenges that emerge when such platforms integrate vertically across various stages of their product and service production and distribution. Through four empirical case studies, we demonstrate how combining large-scale analysis with consumer surveys can uncover issues such as preferential treatment by platforms and discrepancies between platform offerings and consumer preferences.

We want to highlight four key findings from our research. First, our empirical case studies show that evaluating self-preferencing on digital platforms demands a detailed, case-by-case analysis. These platforms use a variety of complex choice architectures, each of which may include self-preferencing elements. A thorough analysis is essential for regulators and researchers to identify such elements, a task made difficult by the often opaque nature of platform design decisions. Distinguishing between beneficial and detrimental effects of choice architectures can be challenging, and it remains to be seen whether data access rights and reporting duties, as established under the European Union's Digital Markets Act and Digital Services Act, will be sufficient to overcome this challenge.

Second, our empirical case studies demonstrate how comparing choice architectures with consumer preferences can allow for insights into which choice architectures may be problematic. They also showcase how surveys can provide an avenue for comparing the impact of product offering choice architectures with consumer preferences. For instance, if a choice architecture nudges consumers towards product A over product B, but surveys show a clear preference for product B, this discrepancy could point towards a mismatch between choice architectures and consumer preferences. 

It is important to stress that we do not view consumer surveys as flawless or the sole method for understanding consumer preferences. Browser extensions and click-stream data present alternative methods for capturing consumer preferences, enabling researchers and regulators to directly observe consumer search and purchasing activities. Consumer panels, often equipped with custom browser extensions, have previously been employed for algorithmic auditing, particularly to observe search results and product features \citep{farronato2023}. Observing actual consumer behavior through click-stream data could be another avenue to explore consumer preferences \citep{bechtold2014}. While a detailed exploration of these tools is beyond this article's scope, they could  enable precise comparisons between choice architectures and consumer preferences on digital platforms. As such, these tools could provide an important ingredient in regulators' enforcement actions against digital platforms. 

Empirical investigations such as those explored in this article are no substitute for a thorough legal analysis of platform behavior under applicable legal standards. They should be considered as one component of a broader enforcement strategy, potentially relevant to a range of legal issues. In this article, we deliberately avoid making definitive statements about legal compliance, as a comprehensive legal analysis would encompass factors beyond the scope of the tools and methods discussed here. 

Third, as Section \ref{sec:preftreatment} has shown, choice architectures influence consumer behavior in multiple ways. How product searches are processed and search results presented can influence not only consumer purchasing decisions. It can also shape consumer preferences from the outset. With digital platforms becoming an integral part of our daily lives, establishing a neutral ``ground truth'' -- determining what a consumer would choose in the absence of these architectures -- poses a significant challenge. This may complicate future regulatory efforts that seek to base interventions on a deviation of a platform architecture from an objectively defined ``ground truth.'' For example, uncertainty over whether consumer preferences are influenced by a specific choice architecture could present regulators with difficult questions when contemplating interventions against platforms for perceived self-preferencing practices that seem contrary to consumers' best interests.

Fourth, our case studies on Amazon's choice architectures underscore the need for caution in addressing digital platform issues solely through ex-ante regulation. Regulating choice architectures involves intervening in the core functions of digital platforms, a move that could blur the lines between competition policy and market design. This does not mean that ex-ante regulation is always inappropriate for digital markets. However, it comes with its own set of challenges. In some instances, fine-grained regulatory interventions may become so costly in terms of design, audit, and compliance that a regulator should refrain from such interventions.

On a more general level, our empirical case studies suggest a future where algorithmic auditing becomes a key part of regulating the digital economy. By employing algorithmic auditing tools, regulators and researchers could uncover new forms of self-preferencing previously undetectable without the aid of automated, large-scale analysis. Although our focus here has been on Amazon's choice architecture and its impact on consumers, similar methodologies could investigate its effects on independent sellers, delivery logistics providers, and fulfillment centers. Importantly, the tools discussed in this article have implications beyond regulatory and research interests. Digital platforms themselves could leverage these tools to proactively identify and address potential legal and societal risks they might otherwise inadvertently contribute to. 

Moving towards a future of algorithmic auditing and compliance requires regulations that are relatively easy to audit in an algorithmic, large-scale manner. In such a future, it may be particularly important for policy makers to design regulations in a way that allows for easy algorithmic auditing \citep{sandvig2014auditing,metaxa2021}. This might involve not just granting data access rights to regulators and researchers and adopting standardized data formats, but also designing regulatory obligations that are straightforward to encode and monitor algorithmically. Additionally, clarifying the legal permissions for researchers to use web scraping and other tools for compliance assessment could significantly enhance the effectiveness of these regulations.

%% file: 6_Conclusion.tex
In this article, we have explored how digital platforms whose architecture are characterized by algorithmic systems and deliberate choice architectures can be investigated utilizing large-scale algorithmic audits and consumer surveys. While our empirical case studies focused on Amazon, our approach could be applied to other online platforms that act as an intermediary among various stakeholders as well. We believe that our case studies demonstrate how empirical methods originating in computer science can equip policymakers, regulators, and researchers with the tools necessary for detailed monitoring of digital platform operations. This, in turn, provides valuable empirical evidence to inform the ongoing debate on effective digital market regulation. Additionally, our analysis highlights the difficulties researchers and regulators face when trying to identify anti-competitive behaviors or impacts from observed platform conduct. Specifically, our case studies shed light on the significant challenges in determining whether a platform's choice architecture is designed to unfairly benefit its own interests or to genuinely enhance service quality.

In a future world of regulating digital markets through algorithmic auditing, we envision tools enabling computer scientists, economists and legal scholars to analyze firm behavior under digital regulations, to empirically identify unintended consequences of such regulations, and to predict, \textit{ex ante}, the effects of further or alternative regulatory interventions. We envision tools that, at least partially, automate the auditing and compliance processes, thereby reducing compliance costs for firms while offering regulators automated enforcement mechanisms. Such tools could find applicability beyond antitrust law, extending to privacy law (see \citealt{manandhar2024,zac2024}), content moderation, intellectual property law, and many other legal domains. Regulating digital markets through algorithmic auditing could not be accomplished by one or two of the disciplines on their own. We need computer scientists to develop new tools for large-scale data collection and automated regulation enforcement. We need economists to build frameworks for understanding social implications and welfare effects, alongside econometric methods for measuring causal impacts of firm behavior. And we need legal scholars to study the normative goals of particular regulations, the trade-offs involved in achieving such goals, and to understand the institutional specifics of enforcement mechanisms. In this sense, the answer \textit{to} the machine may not be \textit{in} the machine \citep{clark1996}. It may be in more interdisciplinary research.